\documentclass[aps,prl,twocolumn]{revtex4}

\usepackage{graphicx}
\usepackage{amsmath}

\DeclareMathOperator{\Tr}{Tr}
\renewcommand{\d}{\mathrm{d}}

\renewcommand{\vec}[1]{\mathbf{#1}}
\newcommand{\mr}[1]{\mathrm{#1}}

\newcommand{\mat}[1]{\mathbf{#1}}

\newcommand{\tempnote}[1]%
   {\begingroup{\it (NOTE: #1)}\endgroup}

\begin{document}

\title{Charge transport in ballistic multiprobe graphene structures}

\author{M.~A. Laakso}
\author{T.~T. Heikkil\"a}
\email[]{Tero.Heikkila@tkk.fi}
\affiliation{Low Temperature
Laboratory, Helsinki University of Technology, P.O. Box 5100
FIN-02015 TKK, Finland}

\date{\today}

\begin{abstract}
We study the the transport properties of multiterminal ballistic graphene samples, concentrating on the conductance matrix, fluctuations and cross-correlations. Far away from Dirac point, the current is carried mostly by propagating modes and the results can be explained with the conventional semiclassical picture familiar from ray optics, where electrons propagate along a single direction before scattering or reaching the terminals. However, close to the Dirac point the transport is due to evanescent modes which do not have to follow a rectilinear path. As we show in this Letter, this property of the evanescent modes influences the conductance matrix. However, at best it can be observed by measuring the cross correlations in an exchange Hanbury Brown-Twiss experiment.
\end{abstract}


\maketitle

%\section{Introduction}
Transport in undoped graphene is entirely due to evanescent modes and can be seen as a form of electron tunneling. In contrast, lightly doped graphene supports propagating modes that have a linear dispersion relation, similar to the $\vec{k}\cdot\vec{P}$ approximation of semiclassical electron dynamics. Semiclassical electrons can be described in terms of classical trajectories and a picture of ray optics, whereas it is not as clear how evanescent waves move in a sheet of graphene. One way to provide more insight to this duality of evanescent and propagating modes is the study of cross-conductances and cross-correlations of electric current in multiprobe graphene structures. This is the aim of our present Letter.

One of the striking properties of electronic transport due to evanescent modes in graphene is the ``pseudodiffusive'' behaviour in undoped samples, manifesting itself in the transmission distribution of ballistic graphene at the Dirac point \cite{tworzydlo_prl}. It turns out that all the different cumulants of current (fluctuations) through such a graphene sample behave in the same way as they would in a diffusive wire \cite{beenakker_rmp}. However, the nature of conduction in the two cases is quite different: in diffusive wires, the conduction electrons are propagating but due to multiple scattering the information on the propagation direction in such systems is (almost) lost. One outcome of this momentum isotropization can be observed in an exchange Hanbury Brown-Twiss cross-correlation experiment, where the measurement result depends on coherent processes that connect all terminals involved in the measurement \cite{blanter+buttiker}. Making such a cross-correlation experiment in conventional ballistic conductors would yield a vanishing result, because in the absence of scattering propagating modes can only couple pairs of terminals.

\begin{figure}
 \centering
 \includegraphics[width=7.8cm]{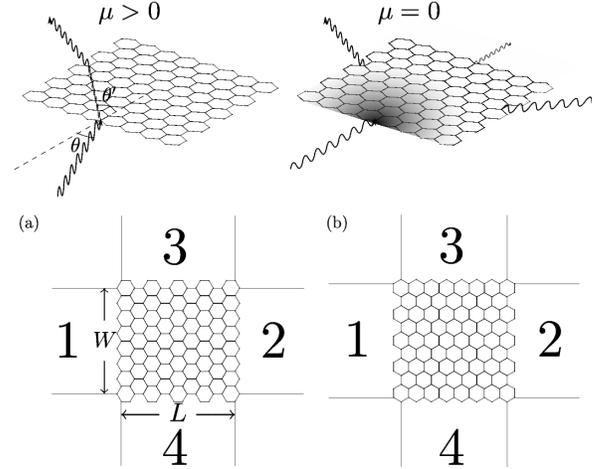}
 \caption{(Top) A schematic explanation for the difference between ray optics and evanescent optics picture of electron propagation. Propagating modes (left) couple only two of the leads in the absence of elastic scattering. Which lead the electron enters depends on the angle of the electron beam at the first lead-graphene contact. This fact can be used to estimate the dependence of the cross-conductances on the spatial dimensions of the scattering region (see Eq.~\eqref{eq:rayoptics}). Evanescent modes (right) in contrast couple all of the leads. (Bottom) Two alternative geometries for the four-probe setup considered in this Letter. In setup (a) interfaces between graphene and leads 1 and 2 are of zigzag type and interfaces to leads 3 and 4 are of armchair type. In (b) graphene sheet has been rotated by 90 degrees, so that the interface types are switched.}
 \label{fig:schematics}
\end{figure}

\begin{figure}
 \centering
 \includegraphics[width=8cm]{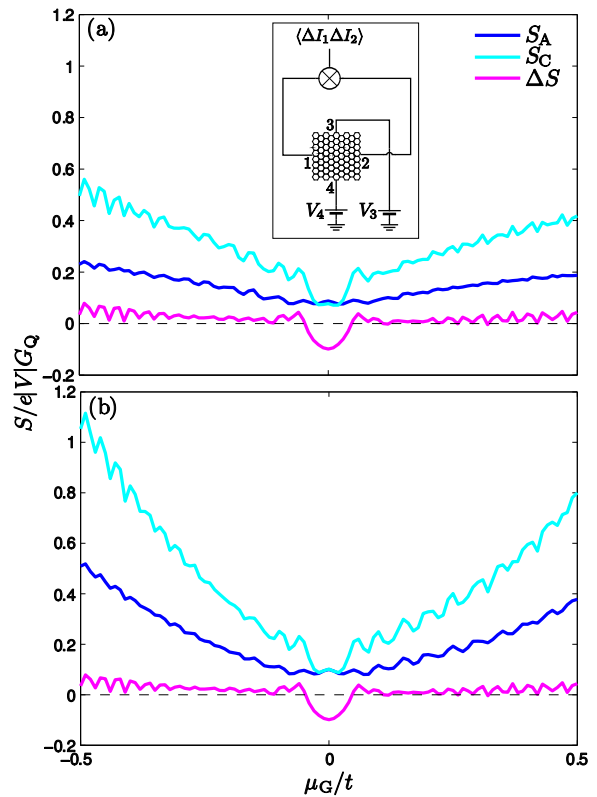}
 \caption{(color online) Noise cross-correlation $S=-S_{12}$ between leads 1 and 2 as a function of chemical potential in graphene $\mu_\mr{G}$. (a) and (b) correspond respectively to setups (a) and (b) in Fig.~\ref{fig:schematics}. Inset shows the biasing scheme for the HBT experiment.}
 \label{fig:hbt}
\end{figure}

In ballistic graphene, there are no elastic scatterers, but due to the evanescent nature of the charge carriers, the momentum direction of the electron waves is not well defined. Rather, the waves are spread out and a single evanescent wave can couple each of the terminals. As a result, the exchange Hanbury Brown-Twiss result will be very similar to that in diffusive wires (see Fig.~\ref{fig:hbt}). The behavior of the two kinds of modes is schematically illustrated in Fig.~\ref{fig:schematics}.

% Electronic properties of graphene have been the subject of extensive theoretical research in the past few years. After numerous experimental results on the conductivity of graphene,\cite{novoselov_nature,kim} the first results on the noise properties have also surfaced.\cite{dicarlo,danneau} The logical continuation of this work would be the study of cross-correlations in graphene.

%In this Article we study theoretically electronic transport in
%multiprobe graphene structures. We calculate cross-conductances,
%Fano factors and cross-correlations in a cross-geometry of four
%terminals. We also study the so called exchange Hanbury Brown-Twiss
%effect. We find that due to the exchange effect in clean graphene,
%noise power is practically unchanged when current is injected from
%two terminals instead of just one -- a similar effect that appears
%in some diffusive systems.

%The structure of this paper is the following: We first briefly
%describe our theoretical framework based on the tight-binding
%approach and scattering formalism. The numerical results and their
%analysis is presented are presented in Sec.~\ref{sec:results} and
%discussed in Sec.~\ref{sec:discussion}.

%\section{Tight-binding approach}\label{sec:tb}
To compute the conductances and correlators, we employ the tight-binding approach, which has been successfully used to describe the transport in graphene \cite{robinson+schomerus,schomerus_prb,blanter+martin} and is also the starting point in the derivation of the Dirac equation describing charge carriers in graphene \cite{peres}. All the numerical results of this Letter are obtained with this approach. On top of this, our aim is to describe how well these results can be understood on a qualitative level.

In our numerics, the tight-binding Hamiltonian matrix is used to obtain the retarded Green's function of the scattering region via the equation
\begin{equation}
 (E+i\eta-\mat{H}-\mat{\Sigma})\mat{G}^R=\mat{I},
\end{equation}
where the coupling to the leads is described by a self-energy $\mat{\Sigma}$. It is then a straightforward task to obtain the scattering matrix from the Fisher-Lee formula \cite{fisher+lee}:
\begin{equation}\label{eq:smatrix}
 \mat{s}=-\mat{I}+i\sqrt{\mat{v}}\mat{\Psi}^\dagger\mat{P}^\dagger\mat{G}^R\mat{P}\mat{\Psi}\sqrt{\mat{v}}.
\end{equation}
Here $\mat{v}$ is a diagonal matrix containing the propagation velocities of the modes, $\Psi_{mn}=\sqrt{\frac{2}{W+1}}\sin\frac{mn\pi}{W+1}$, where $m$ numbers the sites in the leads and $n$ is the mode index, and $\mr{P}_{mn}$ is a matrix that contains the hopping amplitudes from graphene site $m$ to lead site $n$.

Once the scattering matrix has been found, we use it to calculate the linear conductance \cite{buttiker92}
\begin{equation}
G_{\alpha\neq \beta}=\left.\frac{\d I_\alpha}{\d
V_\beta}\right|_{V=0}=G_Q
\Tr\left[\mat{s}^{\alpha\beta\dagger}\mat{s}^{\alpha\beta}\right]
\label{eq:conductance}
\end{equation}
and the zero-frequency cross correlators \cite{blanter+buttiker}
%Zero-frequency
%cross-correlations between two leads are given
%by\cite{blanter+buttiker}
%\begin{align}
% S_{\alpha\beta}&=\frac{e^2}{2\pi}\sum_{\gamma\delta}\int\d E\Tr\left[\mat{A}^{\gamma\delta}(\alpha,E,E)\mat{A}^{\delta\gamma}(\beta,E,E)\right] \nonumber \\
% \times&\left\{f_\gamma(E)(1-f_\delta(E))+(1-f_\gamma(E))f_\delta(E)\right\},
%\end{align}
%where
%\begin{equation}
% \mr{A}^{\gamma\delta}_{mn}(\alpha,E,E')=\delta_{mn}\delta_{\gamma\alpha}\delta_{\delta\alpha}-\sum_j\left(\mr{s}^{\alpha\gamma}_{jm}\right)^\ast(E)\mr{s}^{\alpha\delta}_{jn}(E').
%\end{equation}
%In the linear response regime we can neglect the energy dependence
%of the scattering matrices and instead evaluate them at the Fermi
%energy. At zero temperature we furthermore have $\int\d E\left\{
%f_\gamma(E)(1-f_\delta(E))+(1-f_\gamma(E))f_\delta(E)\right\}=|\mu_\delta-\mu_\gamma|$
%so that
\begin{equation}\label{eq:noise}
 S_{\alpha\beta}=G_Q\sum_{\gamma\neq\delta}\Tr\left[\mat{s}^{\alpha\gamma\dagger}\mat{s}^{\alpha\delta}\mat{s}^{\beta\delta\dagger}\mat{s}^{\beta\gamma}\right]|\mu_\delta-\mu_\gamma|.
\end{equation}
Here $G_Q=2e^2/h$ and $\mu_i$ are the potentials in the leads. For noninteracting fermions, cross-correlations between different leads, $\alpha\neq\beta$, are always negative \cite{blanter+buttiker}. Note that we defined the conductance such that it corresponds to linear response of the current $I_\alpha$ in lead $\alpha$ when the potential $\mu_\beta=-eV_\beta$ in lead $\beta$ is slightly varied and the potentials of all other leads are kept constant.

According to Eq.~\eqref{eq:noise}, the cross-correlators are dependent on both direct processes coupling only the two leads (when $\gamma$ and $\delta$ are both either $\alpha$ or $\beta$) where the correlators are measured, or indirect processes involving also other leads. These two types of contributions can be separated in an exchange Hanbury Brown-Twiss experiment \cite{blanter+buttiker,buttiker_hbt}. There the noise generated by one current source alone is compared to the noise generated by two current sources. Noise correlations between leads 1 and 2, defined in the following by $S=-S_{12}$, are studied in three different cases: In experiment A voltage $V$ is applied to reservoir 3, whereas in experiment B voltage $V$ is applied to reservoir 4. In experiment C voltage $V$ is applied to both reservoirs. The biasing scheme is shown in the inset of Fig.~\ref{fig:hbt}. Classically, $S_C=S_A+S_B$, but quantum mechanical interference effects give rise to an exchange term, $\Delta S=S_C-S_A-S_B$. The latter is given by \cite{blanter+buttiker}
\begin{equation}
\Delta S = \frac{4e^3|V|}{h}\Tr\left[\mat{s}^{14\dagger}
\mat{s}^{13}\mat{s}^{23\dagger}\mat{s}^{24}+\mat{s}^{13\dagger}\mat{s}^{14}\mat{s}^{24\dagger}\mat{s}^{23}\right].
\end{equation}
In general, the exchange term can have either sign. Negative $\Delta S$ means that exchange effects suppress noise whereas positive $\Delta S$ implies enhanced noise. Moreover, a finite $\Delta S$ can be present only if there are modes that couple simultaneously all four leads so that the corresponding elements for all the scattering matrices above are non-zero. As we show below, the latter is the case in our system for evanescent modes close to the Dirac point, but not for the propagating modes far away from it \cite{trajectorynote}.

%\section{Results}\label{sec:results}
% \tempnote{This paragraph may not be so necessary --- we have anyway a very long introduction!} We apply the multiprobe scattering formalism to a cross-geometry formed of a graphene sheet at the intersection of four normal metallic leads. We first study the cross-conductances between different leads and their dependence on sheet dimensions and chemical potential in graphene. We also analyze our results far away from the Dirac point using a simple ray-optics picture, which describes part of the results. After this we turn onto investigating the correlations. As we show in the numerics, the autocorrelations, i.e., the shot noise can be described with a single Fano factor both in the case of local and non-local noise. The final part of our work concentrates on the exchange Hanbury Brown-Twiss experiment and it confirms the picture of evanescent waves coupling each of the terminals.

We apply the multiprobe scattering formalism to a cross-geometry formed of a graphene sheet at the intersection of four metallic leads. In our practical implementation we use a graphene sheet with some 8500 lattice points (unless specified otherwise).
%This is large enough to suppress most finite-size effects.
The square lattice leads are matched to the graphene sheet so that the lattice constant in the leads connected to a zigzag edge is $a_\mr{L}=a$ and in the leads connected to an armchair edge $a_\mr{L}=a/\sqrt{3}$, where $a\approx2.46\:\text{\AA}$ is the lattice constant in graphene. The Fermi level in the leads is chosen to correspond to half-filling of the band. This allows for a description of good contacts \cite{schomerus_prb}.

Numerical simulations such as ours are often prone to effects related to the finite size of the simulated lattice. Such finite-size effects are relevant in the study of nanoribbons \cite{han_prl}, but in typical experimental samples with dimensions in excess of 100 nm these effects are washed out. In our simulations, these finite-size effects are related to the change of the number of propagating modes inside graphene, and therefore lead to rapid oscillations of the calculated quantities vs.~gate potential. Moreover, in graphene with zigzag edges there forms an edge state \cite{brey+fertig} which behaves differently from the rest of the states. In most experimental systems the effect of this state is fairly small, but it affects some of our numerical results.
%\tempnote{I am not 100-\% sure of this --- with this I was referring to the fact that the skew conductances depend on the size of the lattice}
In our numerics, two of the edges are always of the zigzag type and two of them are of the armchair type. To take into account the effect of the type of the edges at the graphene--lead interface, we consider two alternative geometries, depicted in Figs.~\ref{fig:schematics}(a) and (b).

\begin{figure}
 \centering
 \includegraphics[width=8cm]{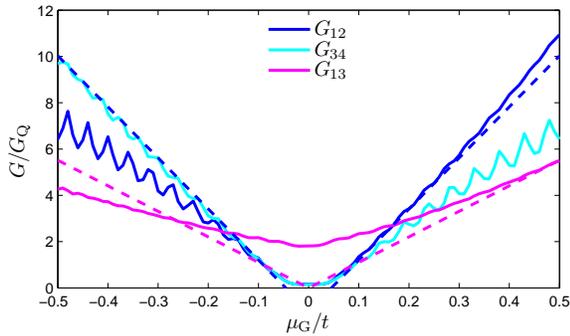}
 \caption{(color online) Cross-conductances in setup (a) of Fig.~\ref{fig:schematics} as a function of chemical potential $\mu_\mr{G}$ in graphene. Dashed lines show the predictions from the ray-optics picture, Eq.~\eqref{eq:rayoptics}, shifted by a constant to take into account the contribution of the evanescent modes. Note that the slope $\d G/\d\mu_\mr{G}$ of these predictions is roughly in accord with the numerical results.}
 \label{fig:crosscond}
\end{figure}

%\subsection{Cross-conductances}\label{sec:crosscond}
The cross-conductances for a square sheet of graphene as a function of the chemical potential $\mu_\mr{G}$ are shown in Fig.~\ref{fig:crosscond}. The direct conductances obey the approximate symmetry $G_{12}(+\mu_\mr{G})\approx G_{34}(-\mu_\mr{G})$ and are quite small at the Dirac point $\mu_\mr{G}=0$. The ``skew'' conductance $G_{13}$ is much larger at the Dirac point. It increases with increasing $|\mu_\mr{G}|$ as well, but with a slope that is roughly half of that in the direct conductance. The evanescent modes are highly localized at the graphene--lead interfaces and therefore the coupling between adjacent leads is strong at the Dirac point, leading to a relatively high $G_{13}$.

\begin{figure}
 \centering
 \includegraphics[width=\columnwidth]{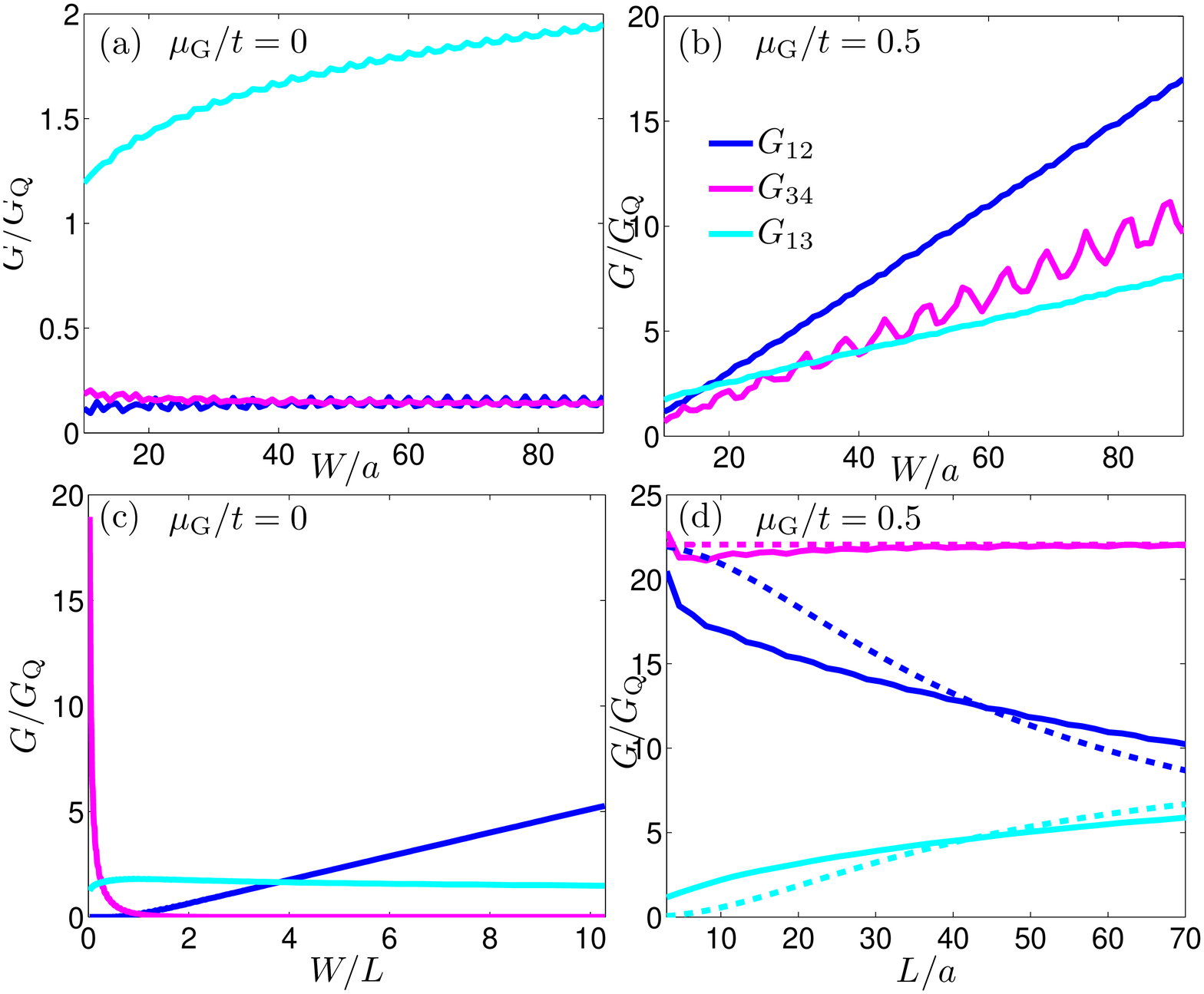}
 \caption{(color online) Cross-conductances as a function of the size of the graphene sheet. In (a) and (b), the aspect ratio $W/L=1$ is kept constant while the number of lattice points is increased. (a) is plotted at the Dirac point (evanescent modes) and (b) for $\mu_\mr{G}/t=0.5$ (propagating modes). In (c) and (d) we plot the cross-conductances as a function of the aspect ratio at the Dirac point and for $\mu_\mr{G}/t=0.5$, respectively. In (c) the graphene sheet had a fixed size of $WL=3600a^2$ and in (d) we chose $W=60a$. In (d) the dashed lines show the prediction from Eq.~\eqref{eq:rayoptics}.}
 \label{fig:xcondsize}
\end{figure}

We have also studied the geometry dependence of the cross-conductances by varying the size and aspect ratio of the graphene sheet. Figures \ref{fig:xcondsize}(a) and (b) show the cross-conductances as a function of the size of the graphene sheet. The aspect ratio was kept as close to unity as possible. For the evanescent modes (a) the direct conductances are constant, $G\approx0.15G_Q$, as can be expected from the pseudodiffusive model. The skew conductance increases with size, and approaches asymptotically the value $G_{13}=2G_\mr{Q}$, one conductance quantum for graphene with spin and valley degeneracies. The deviation of $G_{13}$ from $2G_\mr{Q}$ is thus a finite-size effect, presumably caused by the quasibound state near the zigzag edges \cite{brey+fertig}. For the propagating modes (b) all conductances increase linearly with size, in line with the linear increase in the number of modes.

Fig.~\ref{fig:xcondsize}(c) shows the cross-conductances as a function of the aspect ratio of the graphene sheet for $\mu_\mr{G}/t=0$. The size of the graphene sheet was kept constant. The direct conductances scale as $G_{12}\propto W/L$ and $G_{34}\propto L/W$ as expected from Ref.~\onlinecite{tworzydlo_prl} with a conductivity of $4e^2/\pi h$. The skew conductance is quite insensitive to the aspect ratio, deviating from its bulk value only at very small or large aspect ratios where finite-size effects play a greater role.

%\begin{figure}
% \centering
% \includegraphics[width=8cm]{xcondratioe.eps}
% \caption{(color online) Cross-conductances as a function of the aspect ratio of the graphene sheet for a fixed size of $WL=3600a^2$ at the Dirac point.}
% \label{fig:xcondratioe}
%\end{figure}
For large $\mu_\mr{G}$ we can assume that all modes are propagating. In this case we can estimate the relative magnitudes of the direct and skew conductances with the ray optics model: In the ballistic limit all modes are ideally transmitted, and the terminal from which the electron exits is determined by the angle of propagation, $k_F\sin\theta=k_y$, where $k_F=2\mu_\mr{G}/(\sqrt{3}ta)$. For large structures we can convert the summation over transverse modes in the Landauer formula to an integral and obtain
\begin{align}\label{eq:rayoptics}
 \frac{G_{12}}{G_\mr{Q}}&=\frac{2Wk_F}{\pi}\frac{W/L}{\sqrt{4+(W/L)^2}},\nonumber \\ \frac{G_{13}}{G_\mr{Q}}&=\frac{Wk_F}{\pi}\left(1-\frac{W/L}{\sqrt{4+(W/L)^2}}\right).
\end{align}
%However, a sizable portion, $N_e/N_p\approx(\sqrt{3}\pi-2\mu_G /t)/(2\mu_G/t)$ of the modes are evanescent and couple predominantly to the neighboring terminals.
Figure \ref{fig:xcondsize}(d) shows the cross-conductances as a function of the length of the graphene sheet for $\mu_\mr{G}/t=0.5$. The sum of all cross-conductances is constant for large samples and agrees with Eq.~\eqref{eq:rayoptics}, confirming that all propagating modes are ideally transmitting. This also implies that evanescent modes, not taken into account in Eq.~\eqref{eq:rayoptics}, do not contribute to conductance. The individual cross-conductances deviate somewhat from the raytracing model, however, but the slope seems to be correct for large samples. This deviation is probably due to the relatively small size of our structure, where the replacement of the sum by an integral is not fully justified.
%\begin{figure}
% \centering
% \includegraphics[width=8cm]{xcondratiop.eps}
% \caption{(color online) Cross-conductances as a function of the length of the graphene sheet for a fixed width of $W=60a$ at $\mu_\mr{G}/t=0.5$. Solid lines are the numerical results and dashed lines the corresponding analytical approximations with Eq.~\eqref{eq:rayoptics}.}
% \label{fig:xcondratiop}
%\end{figure}

%\subsection{Fano factors}\label{sec:fano}
We now turn to the noise correlations in the four probe setup. We first assume that current is driven between terminals 1 and 2 and the potential in terminals 3 and 4 is kept floating so that no average current flows in them. Due to symmetry the floating potential lies halfway between the potentials in terminals 1 and 2.

\begin{figure}
 \centering
 \includegraphics[width=8cm]{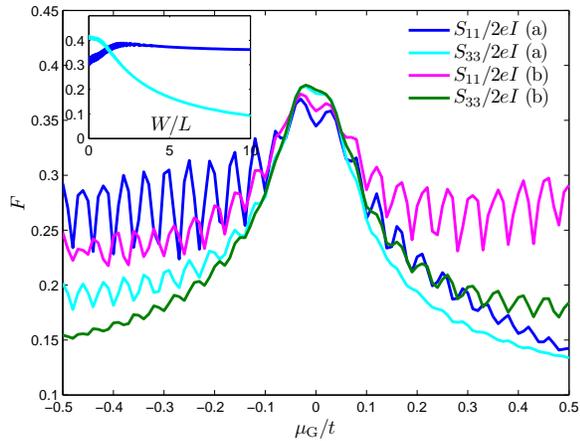}
 \caption{(color online) Local and non-local Fano factors as a function of the chemical potential in graphene. Lines with (a) or (b) in the legend refer to the corresponding setups in Fig.~\ref{fig:schematics}.
 Inset shows the dependence of the Fano factors vs. aspect ratio $W/L$ at the Dirac point.}
 \label{fig:fano}
\end{figure}

We define the local Fano factor with $F_\mr{local}=S_{11}/2eI$, where $S_{11}$ is the noise autocorrelator (shot noise) in terminal 1 and $I$ the average current flowing from 1 to 2. Similarly, the nonlocal Fano factor is defined with $F_\mr{nonlocal}=S_{33}/2eI$. Symmetry again dictates that $S_{11}=S_{22}$ and $S_{33}=S_{44}$. These Fano factors are shown in Fig.~\ref{fig:fano} for both setups (a) and (b) as a function of the chemical potential in graphene with $W=L$. Each of them exhibit a peak of $F\approx0.37$ at the Dirac point and decrease with increased doping of the graphene sheet. The oscillations at finite chemical potentials signal the appearance of propagating modes in the graphene sheet and their period is of the order of level spacing. The nonlocal Fano factors decrease more rapidly and exhibit weaker oscillations. Shot noise also obeys roughly the symmetry $F^{(a)}(\mu_\mr{G})=F^{(b)}(-\mu_\mr{G})$, similar to the cross-conductances.
Upon increasing the value of $W/L$ the local Fano factor of undoped graphene stays unchanged in both setups, but the nonlocal Fano factor decreases, reaching $0.1$ at $W/L=10$ (see inset of Fig.~\ref{fig:fano}). The deviation of $F_\mr{local}$ from 1/3, predicted in \cite{tworzydlo_prl} and observed in \cite{danneau} for two-terminal samples, is probably due to the effective inelastic scattering caused by the presence of the additional terminals 3 and 4.

%\subsection{Exchange Hanbury Brown-Twiss effect}\label{sec:hbt}
Figures \ref{fig:hbt}(a) and \ref{fig:hbt}(b) show the exchange Hanbury Brown-Twiss cross correlations for setups (a) and (b), respectively. Near the Dirac point exchange correction is negative, and of similar magnitude with the classical contribution. For this reason two-terminal noise is almost identical to one-terminal noise, $S_{12}/e|V|G_\mr{Q}\approx -0.1$. This finding is similar to the case of disordered box \cite{blanter+buttiker_prb} where transport is diffusive. In ballistic graphene, this result is due to the fact that evanescent states couple all the terminals. For larger $\mu_G$ the exchange correction is almost vanishing (up to a one-channel interference effect). The only differences between the two geometries are the magnitudes of the noise cross-correlations at increasing chemical potentials, which in setup (b) grow roughly quadratically versus the linear increase in setup (a). The negative exchange correction near the Dirac point seems to be quite robust, appearing also with non-square sheets and different interface transparencies.

%\section{Discussion}\label{sec:discussion}
In conclusion, we have calculated cross-conductances, Fano factors and noise cross-correlations in graphene by applying a numerical tight-binding model. With these results we have pointed out that the semiclassical ray-optics picture usually valid in large ballistic normal conductors is also valid in doped  graphene, but cannot be used to describe the behavior of the evanescent modes.

\begin{acknowledgments}
This work was supported by the Academy of Finland and the NANOSYSTEMS/Nokia contract with the Nokia Research Center. We thank Izak Snyman, Yaroslav Blanter and Carlo Beenakker for useful discussions. TTH acknowledges the hospitality of the Kavli Institute of Nanotechnology at the Delft University of Technology, where part of this work was carried out.
\end{acknowledgments}

\bibliography{graphene}

\begin{thebibliography}{15}
\bibitem[{tra()}]{trajectorynote}
\bibinfo{note}{For propagating modes, we can use the picture of electron
  trajectories, similar to for example Ref.~\onlinecite{blanter+buttiker}.
  However, this concept is poorly defined for evanescent modes.}
\end{thebibliography}

\end{document}